# Cascaded optical nonlinearities in dielectric metasurfaces

Sylvain D. Gennaro[1,2], Chloe F. Doiron[1,2], Nicholas Karl[1,2], Prasad P. Iyer[1,2], Michael B. Sinclair[1,2] and Igal Brener [1,2*]

*1 Sandia National Laboratories, Albuquerque, New Mexico, USA;*
*2 Center for Integrated Nanotechnologies, Sandia National Laboratories, Albuquerque, NM 87185, USA;*

Since the discovery of the laser, optical parametric nonlinearities have been at the core of efficient light conversion sources. Typically, thick transparent crystals or quasi-phase matched waveguides, are utilized in conjunction with phase-matching techniques to select a single parametric process. In recent years, due to the rapid developments in artificially structured materials, optical frequency mixing has been achieved at the nanoscale in subwavelength resonators arrayed as metasurfaces. Phase matching becomes relaxed for these wavelength-scale structures, and all allowed nonlinear processes can, in principle, occur on an equal footing. This could promote harmonic generation via a cascaded (consisting of several frequency mixing steps) process. However, so far, all reported work on dielectric metasurfaces have assumed frequency mixing from a direct (single step) nonlinear process. In this work, we prove the existence of cascaded second-order optical nonlinearities by analyzing the second and third wave mixing from a highly nonlinear metasurface in conjunction with polarization selection rules and crystal symmetries. We find that the third wave mixing signal from a cascaded process can be of comparable strength to that from conventional third harmonic generation, and that surface nonlinearities are the dominant mechanism that contributes to cascaded second order nonlinearities

In a nonlinear medium, harmonic generation at three times the pump frequency, $\omega$, can be achieved by traditional third-harmonic generation (THG) from a direct $\chi^{(3)}$ process or via a $\chi^{(2)}$:$\chi^{(2)}$ cascaded process of second harmonic generation (SHG) followed by sum frequency generation (SFG) as illustrated in figure 1. [1–5] To distinguish this cascaded process from conventional THG, we denote the cascaded process as $\chi^{(2)}$:$\chi^{(2)}(3\omega)$ and conventional THG as $\chi^{(3)}$:$THG$. Four-wave mixing, one of the hallmarks of third order nonlinearities, can also be achieved by SHG of one pump beam, followed by difference frequency generation (DFG) with a second pump beam. [6] Traditionally, $\chi^{(2)}$:$\chi^{(2)}(3\omega)$ is weak in bulk media since phase-matching limits frequency mixing to one process at a time. [7–9] In nonlinear metasurfaces, due to the relaxation of phase-matching limitations, several of the frequency mixing products observed experimentally might be explained by cascaded second order optical nonlinearities. [10–14] However, given that both $\chi^{(3)}$:$THG$ and $\chi^{(2)}$:$\chi^{(2)}(3\omega)$ occur at the same frequencies, and are assumed to have the same polarization response in cubic crystal systems, [15] it is difficult to distinguish between the two processes in these types of non-phase matched systems. Understanding the origin of light conversion in these systems is critical since the efficiencies of cascaded harmonic generation could compete with those of conventional direct frequency mixing processes or even surpass them. [16,17]

In this work, we prove the existence of cascaded second order optical nonlinearities by investigating two and three wave mixing from a dielectric metasurface made of GaAs nanocylinder resonators, isolated from a GaAs/AlGaAs substrate by a AlGaO spacer layer (see the methods section for additional information on the fabrication process).

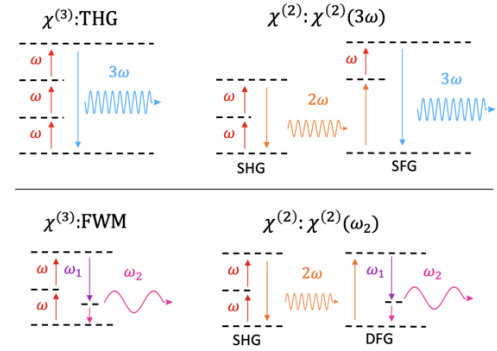

**Figure 1. Cascaded versus direct frequency mixing processes.** Schematic of harmonic generation of a wave: at frequencies $3\omega$ via direct third harmonic generation $\chi^{(3)}$: $[\chi^{(3)}(-3\omega;\omega,\omega,\omega)]$ or through cascaded second harmonic generation (SHG) and sum frequency generation (SFG). $[\chi^{(2)}(-2\omega;\omega,\omega):\chi^{(2)}(-3\omega;2\omega,\omega)]$; and at frequencies $\omega_2$ via direct $\chi^{(3)}$ − FWM $[\chi^{(3)}(-\omega_2;\omega_1,\omega,\omega)]$ or through cascaded SHG and DFG $[\chi^{(2)}(-2\omega;\omega,\omega):\chi^{(2)}(-\omega_2;\omega_1,2\omega)]$.

The metasurface is illustrated in figure 2a. We choose this design as it possesses a symmetric resonator geometry and the metasurface is known to exhibit intense harmonic emission at multiples of the pump frequency enhanced by the excitation of Mie-like optical resonant modes at the fundamental (i.e. pump) frequency (see fig. 2b). [18] The linear reflectance spectra shows a broad resonance with two distinct peaks characteristic of two Mie-like optical modes, namely an electric dipole (ED) mode centered around 1200nm and a magnetic dipole (MD) mode centered around 1400nm. [14,18] Additional electric field data of the electric and magnetic dipole modes are shown in supplementary figure S1.At pump wavelengths resonant with the ED (1200nm) and MD modes (1400nm), the harmonic emissions at $2\omega$ and $3\omega$ are above the bandgap of GaAs and the substrate absorbs the radiated harmonics.

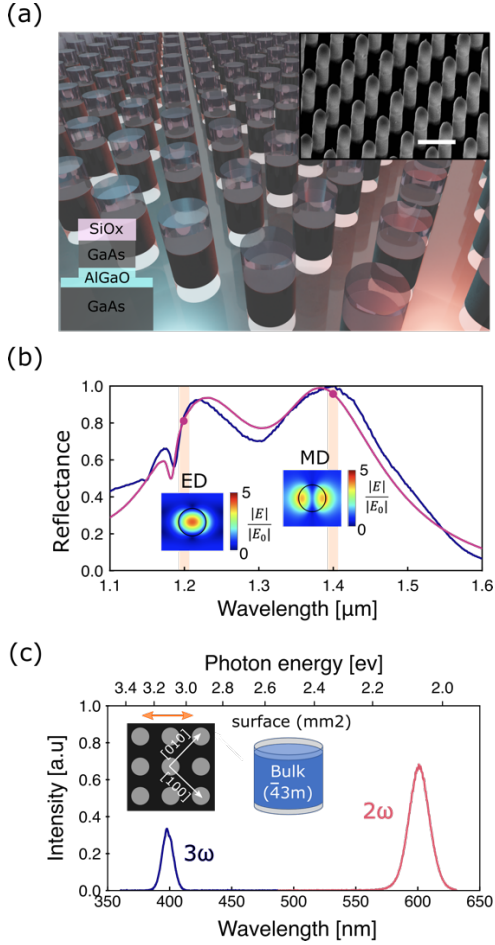

**Figure 2. Optical modes and symmetry of our nonlinear metasurface.** (a) Artistic representation of the dielectric metasurface considered in this work Top inset is a tilted view of a scanning electron micrograph (SEM) of the metasurface. White scale bar represents 1 μm. Bottom inset shows the different materials that constitutes each resonator. (b) Simulated (solid purple) and measured (solid blue) linear reflection spectra of our metasurface. Top views of normalized electric field distribution, $\frac{|E|}{|E_0|}$ of the resonators are displayed for the electric dipole mode at 1200 nm and the magnetic dipole mode at 1400 nm. (c) Measured spectra of the nonlinear signal at $2\omega$ (solid pink) and $3\omega$ (solid blue) excited with a 1200 nm linearly polarized pump laser beam. Insets are a schematic of the metasurface showing the orientation of the GaAs crystal axes [100] and [010] as white arrows, and the incident pump light polarization as a double orange arrow. The second schematic represents a GaAs nanocylinder where the crystal point-group symmetry of bulk ($\bar{4}3m$) and top/bottom surface (mm2) are indicated.

Thus, we collect the harmonic emission using a reflection geometry. Our optical apparatus for the nonlinear measurements is described in the methods section and its schematic is presented in supplementary figure S2. Examples of harmonic spectra at $2\omega$ and $3\omega$ are shown in figure 1c for a pump wavelength of 1200nm, with an incident peak irradiance of 2GW/cm², and incident polarization parallel to the horizontal edge of our nanocylinder array, as illustrated by the orange double-sided arrows in the inset of figure 1c. The crystal axis [100], [010] of GaAs are at -45 and 45 degrees with respect to the edge of our metasurface. Since the second and third order nonlinear susceptibility tensors are described in the principal axis frame of the crystal ([100], [010], [001]), we perform the appropriate coordinate transformation when we subsequently model the nonlinear harmonic signal at $2\omega$ and $3\omega$ in figure 3 (see supplementary note 1). Nonlinear optical modeling is carried out with a commercial finite-element solver including nonlinear optical tensors (see supplementary information). Polarization-dependence modeling is then carried out self-consistently both at $2\omega$ and $3\omega$.

In previous work, a $\chi^{(2)}{:}\chi^{(2)}$ ($3\omega$) process was isolated by rotating the sample in-plane and measuring the anisotropy of the total third harmonic emission that arises from $\chi^{(2)}{:}\chi^{(2)}$ ($3\omega$) interfering with $\chi^{(3)}{:}THG$ [17,19]. Since the GaAs crystal belongs to the point-group symmetry $\bar{4}3m$, $\chi^{(3)}{:}THG$ has a fourfold symmetry with respect to crystal orientation as it is commonly observed for cubic crystals. [17,19–21] We expect this to also be true for our metasurfaces due to the rotational symmetry of the nanocylinders and the fourfold symmetry of the cylinder array. However, in our measurements, we observe distinct deviations from a fourfold pattern which we will demonstrate to originate from a cascaded second order process. To understand this cascaded process, we also investigate the symmetry of the optical fields at $2\omega$ (SHG) and the second order nonlinear susceptibility with respect to crystal orientation.

Figure 3 shows measured and simulated normalized intensity of the harmonic emissions at $2\omega$ and $3\omega$ as a function of incident polarization when optically pumping the electric dipole mode at 1200nm (fig. 3 a, b), and the magnetic dipole mode at 1400nm (fig. 3 c, d). We begin by discussing the nonlinear emission at $2\omega$ (fig 3.a and c), where we observe a two-fold symmetry in the measured total SH intensity with maxima centered at 0 and 180 degrees (relative to the axes of the nanocylinder array) for the electric dipole mode and at 90 and 270 degrees for the magnetic dipole mode. However, the simulated second harmonic considering only bulk second order nonlinearities of GaAs ($\chi^{(2)}_{xzy} = \chi^{(2)}_{yzx} = \chi^{(2)}_{zxy} = 200\ [pm \cdot V^{-1}]$ at $\omega$, and $\chi^{(2)}_{zxy} = 450\ [pm \cdot V^{-1}]$ at $2\omega$ [22]) predicts a four-fold symmetry of the SH emission (fig. 3 b, d) with minima for incident light polarized along the respective crystal axes [100] and [010] of GaAs as observed in past literature. [23] This fourfold symmetry is expected as the second order nonlinear polarizability of bulk GaAs is non-zero only for cross terms, $E_i E_j \{i \neq j\}$ of the electric fields.

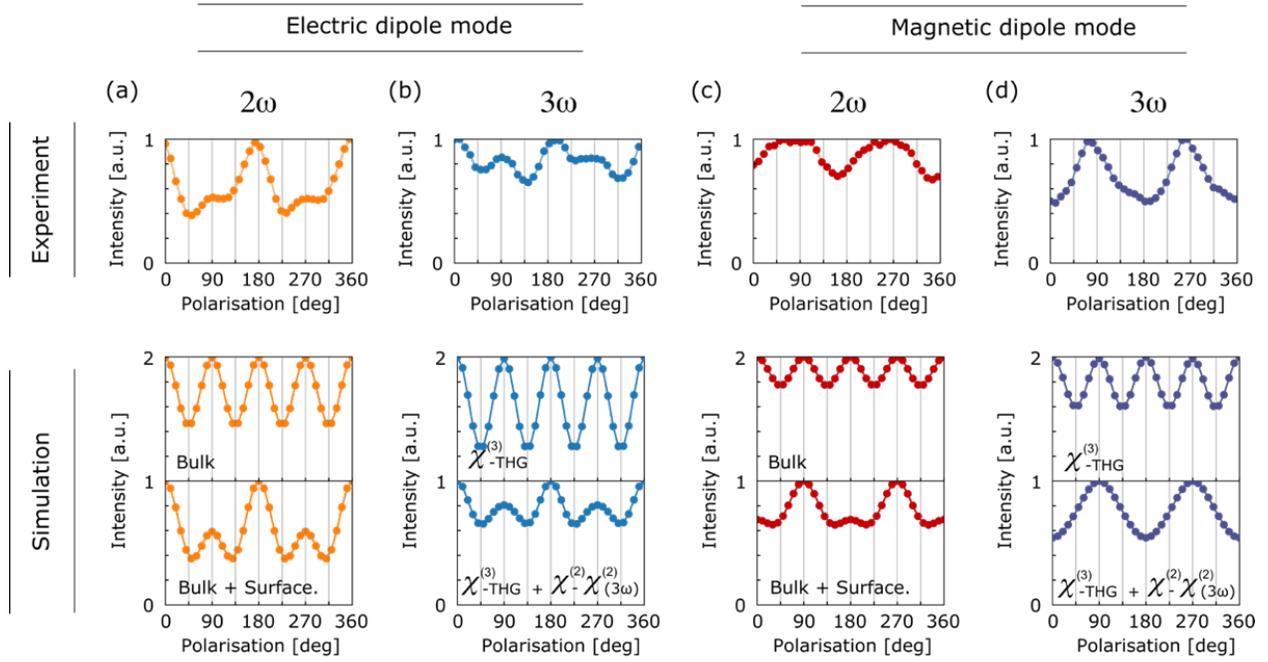

**Figure 3. Identifying cascaded third harmonic with polarization selection rules and crystal symmetry.** Measured (top panel) and simulated (middle and bottom panels) nonlinear harmonic intensities at (a, c) $2\omega$ and (b, d), $3\omega$ and for (a, b) electric dipole mode excitation ($\lambda = 1200\ nm$) and (c, d) magnetic dipole mode excitation ($\lambda = 1400\ nm$) while varying the polarization of the incident beam from 0 to 360 degrees. Fourfold symmetry (middle panels) is characteristic of SHG from bulk nonlinearities and of THG via a direct third order process ($\chi^{(3)}$: THG). Twofold symmetry indicates the presence of SHG from surface nonlinearities at $2\omega$, and of third wave mixing at $3\omega$ via a cascaded second order process. ($\chi^{(2)}$: $(3\omega)$)

The difference between the measured two-fold and predicted four-fold symmetry cannot be explained using bulk second order nonlinearities alone and suggests that surface nonlinearities must play a significant role in the total SH emission. [18,23,24] Indeed, it is well known that the crystal symmetry of the surface of GaAs differs from that of the bulk: GaAs surfaces whose normal vectors are collinear to the principal crystal axes [100], [010], or [001] of GaAs belong to the point-group symmetry mm2. [21,25–27] While surface SHG generally appears as a weak anisotropy in the SHG symmetry studies of unstructured GaAs film, [25–27] dielectric metasurfaces magnify the effect of surface nonlinearities due to the larger total surface area and the strong electric field confinement inside the volume of each nanoscale resonator which leads to greater electric field – surface interaction. [18,23]

Generally, it is challenging to model the resonator's surface nonlinearity on the sidewalls since the normal vector of the surface sidewalls varies with position and the symmetry of the nonlinear tensors are mostly known for surfaces whose normal vectors are aligned with the respective crystal axes of the families [100], [110] or [111]. [21,25,26]. From our experience, we also notice that, during the dry etching process, the surface sidewalls are roughened resulting in amorphous restructuring (see the method section). Thus, the surfaces are no longer crystalline facets with well described local nonlinear polarizations. In contrast, the top and bottom surfaces of the GaAs resonators preserve their epitaxial growth crystallinity, hence their orientation and symmetry remain well defined. Therefore, we consider only the surface nonlinearities at the top and bottom surfaces. Recent work on surface SHG in semiconductors considered regions on the order of a few nanometers at the material/air interface, and demonstrated that this effective volume is a good representation of surface SHG. [24,28] Therefore, in our modeling, we consider two regions at the Air/GaAs and GaAs/AlGaO interface with a thickness of 10nm each and with a symmetry tensor of point-group symmetry mm2. When we include both surface and bulk nonlinearities, with the following surface-induced nonlinear coefficients: $\chi_{zxx}^{(2,s)} = \chi_{zyy}^{(2,s)} = 6\chi_{xzy}^{(2)}$, $\chi_{zzz,}^{(2,s)} = 6\chi_{xzy}^{(2)}$, $\chi_{yxz}^{(2,s)} = \chi_{xzy}^{(2,s)} = 1.8\ \chi_{xzy}^{(2)}$, while maintaining the surface-induced nonlinear tensor symmetry (See supplementary note 1), we successfully retrieve the twofold symmetry in the total SHG intensity (fig3 a, c, bottom panels). Values of the surface nonlinearity coefficients were chosen to reproduce two-fold symmetry in the total second harmonic intensity while respecting the symmetry constraints ($\chi_{zxx}^{(2,s)} = \chi_{zyy}^{(2,s)}, \chi_{zzz,}^{(2,s)}$, $\chi_{yxz}^{(2,s)} = \chi_{xzy}^{(2,s)}$) imposed by the nonlinear tensor for [100] GaAs surfaces. [21] This result reaffirms the importance of surface-induced nonlinear effects in these types of nanophotonic systems.

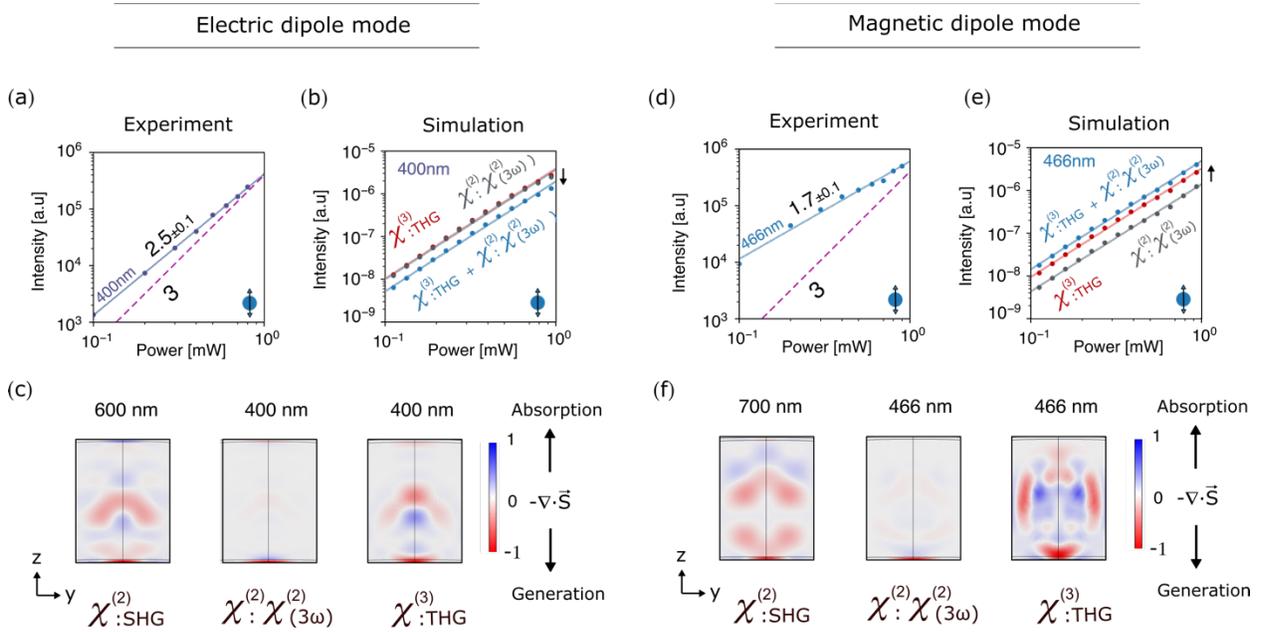

**Figure 4. Origin of cascaded third harmonic in dielectric metasurfaces.** (a, d) Measured nonlinear signal at $3\omega$ (solid dots) as a function of incident power plotted on a logarithmic scale for electric dipole mode excitation at 1200 nm ($3\omega$ @400 nm) and magnetic dipole mode excitation at 1400 nm ($3\omega$ @466 nm) with incident polarisation at 90 degrees. The numeric labels represent the exponent power coefficients, 'a', of each nonlinear process obtained via a linear regression (solid line) on the logarithm of the data such that $\log_{10} P_{out} = a \log_{10} P_{in} + b$ assuming a power fluctuation of our laser of 4%. (b, e) Modelled nonlinear signal at $3\omega$ (solid dots) for $\chi^{(3)}$:THG (solid red dots), $\chi^{(2)}$:$\chi^{(2)}(3\omega)$ (solid grey dots), and the total $3\omega$ signal (solid blue dots). For the ED mode, $\chi^{(2)}$:$\chi^{(2)}(3\omega)$ adds destructively with $\chi^{(3)}$:THG. For the MD mode, $\chi^{(2)}$:$\chi^{(2)}(3\omega)$ adds constructively with $\chi^{(3)}$:THG. Solid lines respectively represent the linear fit for each harmonic process with predicted slopes of (b) 2.6; (e) 2.5. (c) Divergence of the Poyting vector for $\chi^{(2)}$:SHG, $\chi^{(2)}$:$\chi^{(2)}(3\omega)$ and $\chi^{(3)}$:THG. Red and blue areas indicate where harmonic light is generated and dissipated respectively.

We now discuss the polarization dependence of the harmonic emissions measured at $3\omega$. For both the ED and MD resonances, the measured symmetry of the $3\omega$ emission (fig.3 b, d, top panels) deviates significantly from the fourfold symmetry expected from $\chi^{(3)}$:THG when considering bulk third order nonlinearities alone; see fig. 3b, d middle panels. The $\chi^{(3)}(\omega)$, values found to reproduce our third harmonic experimental data: $\chi^{(3)}_{iiii} = 0.52 \times 10^{-19} [m^2 \cdot V^{-2}]$ and $\chi^{(3)}_{iijj} = 0.18 \times 10^{-19} [nm^2 \cdot V^{-2}]$, are within ranges of predicted and measured values for THG. [3,29,30] We note here that surface- $\chi^{(3)}$ processes are negligible when compared to their dipole-allowed, bulk counterparts. [21] To explain this discrepancy in the polarization dependence, we include the cascaded $\chi^{(2)}$:$\chi^{(2)}(3\omega)$ process and consider its interference with $\chi^3$:THG (fig 3b, d, bottom panels). Using this procedure, we can retrieve the twofold symmetry shown in our experiments. When pumping resonantly at the electric dipole mode, the presence of $\chi^{(2)}$:$\chi^{(2)}(3\omega)$ leads to an anisotropy in the total THG and seems to interfere destructively with $\chi^{(3)}$:THG in a similar fashion to past observations in bulk crystalline quartz. [17] For a pump resonant with the magnetic dipole mode however, the harmonic signal measured at $\omega$ shows a more distinct twofold symmetry. For completeness, we have also included in supplementary figure S3 and S4 additional SHG and THG measured and simulated polarization data for some of the other incident wavelengths considered in this work. We find that our model can predict the overall trends of the nonlinear signal at $2\omega$ for other pump wavelengths as the second order nonlinear susceptibilities are relatively constant for the range of wavelengths at $\omega$ and $2\omega$ considered in this work. [22] Since $\chi^{(3)}(\omega)$ values are predicted to vary more significantly over this spectral range, we may need to choose different values of $\chi^{(3)}(\omega)$ to obtain more accurate results for $\chi^{(3)}$:THG at other wavelengths. [3,29] In summary, any twofold symmetry in the harmonic emission at $3\omega$ is a clear signature of a second order cascaded process in our dielectric metasurface.

While characterizing the incident power dependencies of these harmonic processes (see fig. 4a, d), we find that the harmonic signal at $3\omega$ scales sub-cubically for both the magnetic dipole mode ($\approx 1.7$) and the electric dipole mode ($\approx 2.5$). Supplementary figure S6 shows the power measurement of the nonlinear signal at $2\omega$. Sub-power scaling law is attributed in our system to the presence of competing nonlinear processes such as two-photon absorption (2PA) and the subsequent free carrier absorption in GaAs as observed in our previous work. [18] Modelling this effect in our system (see supplementary note S2), we can retrieve sub power scaling laws for both the signal at $2\omega$ and $3\omega$ as shown in figure 4 b and e. While our simplified model predicts correctly the power scaling law for the electric dipole ($\approx 2.5$), it underestimates the effect of 2PA for the magnetic dipole mode given reasonable two photon absorption coefficient for GaAs and field enhancement factors.

Using this self-consistent modeling that is representative of the physics of our metasurface, we can deduce the strength of each harmonic process. We find that, when pumping at the ED mode, $\chi^{(2)}{:}\chi^{(2)}$ ($3\omega$) can be of comparable strength to $\chi^{(3)}{:}THG$ (fig. 4b) and adds destructively to $\chi^{(3)}{:}THG$ such that the total nonlinear signal at $3\omega$ is weakened. However, when pumping the MD mode, we find that $\chi^{(2)}{:}\chi^{(2)}$ ($3\omega$) adds constructively to $\chi^{(3)}{:}THG$ such that the total nonlinear signal at $3\omega$ is strengthened (fig. 4e). Finally, figure 4c and f show the negative divergence of the Poyting vector, $-\nabla \cdot \vec{S}$, which indicates the locations where harmonic light is generated or dissipated. Integrating over the volumes of the resonator that we consider as representative of "bulk" and "surface" effects, we find that $\chi^{(2)}{:}\chi^{(2)}$ ($3\omega$) is primarily generated via second order nonlinearities at the surfaces.

The demonstration of cascaded optical nonlinearities in our dielectric metasurface suggests that alternative pathways to three wave mixing, and more broadly high harmonic generation, [31,32] are achievable at the nanoscale, facilitated by the relaxation of phase-matching. When designing and characterizing novel light conversion sources, it is paramount to account for all parasitic losses and unaccounted frequency mixing mechanisms. In this aspect, our work demonstrates an approach to isolate and characterize cascaded second order optical nonlinearities, based on polarization selection rules and crystal symmetry, and quantify their effect on conventional direct harmonic generation. Our work also reaffirms the importance of considering surface nonlinearities in two wave frequency mixing. Such understanding and analysis could thus be used to significantly enhance efficiencies for frequency-mixing processes as well as their quantum equivalent such as spontaneous parametric down-conversion, and thereby pave the way for a plethora of novel compact light conversion sources based on cascaded second order optical nonlinearities.

**Methods**

**Simulated reflectance spectra.** The reflectance spectra of the metasurface were simulated with finite difference time-domain method (Lumerical Inc) under plane wave excitation with periodic boundary condition. The structure consists of an array of GaAs/AlGaO nanocylinders of diameter 350 nm, periodicity of 750 nm, and a total height of 900 nm. The height of the nanocylinder is partitioned into 300 nm $SiO_x$ etch mask cap (top), 450 nm of GaAs (middle) and 150 nm of AlGaO (bottom). The nanocylinder rests on an estimated 250 nm unetched AlGaO oxide layer on top of a GaAs [001] substrate. The GaAs and $SiO_x$ refractive index value were taken from the database of the finite difference time-domain software. AlGaO was modelled as a non-dispersive medium with refractive index of $n = 1.6$. Additional features such as the reflectance dip observed around 1170 nm arise from off-normal incidence excitation due to the finite numerical aperture of our optical system. This was modelled in simulation as a plane wave tilted by 2.5 deg. Off normal incidence can excite higher order non-radiative modes.

**Reflectance spectroscopy.** The linear reflectance spectra of the metasurface were measured using two home-built white-light spectroscopy setups. The near IR measurement spectra from 900 nm to 1600 nm were measured by focusing the broadband white light from a Stabilized Tungsten-Halogen lamp onto the sample with a 50 mm plano-convex lens. In reflection, the scattering is collected by the same lens and directed into an InGaAs Peltier-cooled spectrometer (NIR Quest 512 from Ocean Optics). Each reflection spectrum is normalized using the reflectance of a gold mirror as a reference measured under the same experimental condition. For the visible regime (400 nm – 900 nm), the incoherent light from a thermal white light source is focused onto our sample with a microscope objective (5X, NA = 0.13). We reduce the back aperture with an iris to limit the range of incidence angles. The light is then redirected onto the entrance slit of an Ocean Optics visible spectrometer. Each reflection spectrum of the metasurface is normalized using the reflectance of a silver mirror as a reference measured under the same experimental condition.

**Sample Fabrication.** The dielectric resonators are fabricated using standard electron-beam lithography and inductively coupled plasma (ICP) dry etch. The sample is grown using molecular beam epitaxy. We start with a GaAs substrate of crystal orientation [001] at the bottom, then three alternating layers of 400 nm thick $Al_{0.85}Ga_{0.15}As$ and 450 nm thick GaAs [001] are grown successively on top (wafer VA0729). For this work, we will only use the top two layers to make the resonators. We then spin coat a negative tone hydrogen silsesquioxane (HSQ Fox-16) electron beam resist. Circular disk patterns are written using 100 keV electron-beam that converts area of exposed HSQ to $SiO_x$. The unexposed resist is developed using tetramethylammonium hydroxide (TMAH) leaving ~300 nm tall $SiO_x$ nanodisks as hard etch masks. We then transfer the pattern onto the GaAs and AlGaAs layers using a chlorine-based (both Cl2 and BCl3) ICP etch. We stop the etch when the AlGaAs layer is etched to half of its depth. Finally, the sample was placed in a tube furnace at ~420 °C for a selective wet oxidization process that converts the $Al_{0.85}Ga_{0.15}As$ into its oxide $(Al_xGa_{1-x})_2O_3$ which has a refractive index of n ≈1.6. Note that the oxide has a larger lattice constant than AlGaAs, which can cause the oxide layer to expand and be slightly larger than 400 nm.

**Nonlinear characterization.** The metasurface is pumped by nominally <80 fs optical pulses at 1 MHz repetition rate with wavelength ranging from 1100 nm to 1600 nm. See supplementary figure S1 for a schematic of the apparatus. The output power is controlled by a pair of a Glan-Taylor polarizer and an achromatic half wave plate. Three long-pass filters with respective cut-off wavelengths of 830 nm, 980 nm, 1064 nm reject any low frequency emission from the laser. Then, the pump beam is redirected onto the back aperture of a microscope lens (20X, NA= 0.45) with a 90:10 (T:R) beam splitter. We place an iris in the beam path to reduce the beam spot diameter from 6 mm to 1 mm, which corresponds to an effective NA of 0.077. Since most of the generated harmonic waves have photon energies above the GaAs bandgap and thus would be absorbed by the GaAs substrate in transmission, we collect harmonic light in reflection by the same objective. The harmonic signal is then redirected with a multimode fiber onto the entrance slit of a spectrometer connected to a Peltier-cooled CCD camera. To control the incident polarization, we place an achromatic 690 – 1200 nm half wave plate in front of the microscope objective. We verify that our optics are polarization insensitive for both the fundamental and harmonic wavelengths and did consecutive measurements to eliminate laser power fluctuations. We note that, in our experiment, our nanocylinders are rotationally symmetric along their main axes, and the linear permittivity of GaAs is isotropic. Therefore, rotating the incident pump polarization at normal incidence is equivalent to rotating the crystal axes (see supplementary figure S3).

**Nonlinear simulation:** Nonlinear simulation were performed with the commercial software COMSOL in the frequency domain using the wave-optic module. First, we calculate the linear electric field at the fundamental frequency, then calculate the respective nonlinear polarizabilities of each process as nonlinear sources. We then extract the average Poyting vector emitted at these nonlinear frequencies from the metasurface. In cubic crystal systems of point-group symmetry $\bar{4}3m$, the third order susceptibility tensor of GaAs has only two independent elements $\chi^{(3)}_{xxxx}$ and $\chi^{(3)}_{xxyy}$ such that the nonlinear polarization of $\chi^{(3)}$: THG can be expressed in the principal-axis coordinate of the crystal, i, [100], j, [010] and k, [001] as $P_i = \epsilon_0[\, 3\chi^{(3)}_{xxyy} E_i(\omega)(E.E) + (\chi^{(3)}_{xxxx} - 3\chi^{(3)}_{xxyy})E_i(\omega)^3, i \in \{x,y,z\}$, [21] where $\epsilon_0$ is the vacuum permittivity, $E_i, E$ are the electric field components at the pump frequency, $\omega$. To model $\chi^{(2)}:\chi^{(2)}(3\omega)$, we first model SHG using second order susceptibility's values of bulk GaAs, and estimated values for surface SHG from our model. Then, we model sum frequency generation using the calculated fundamental and SHG field and the same partition bulk/surface of our nanocylinder. In the principal-axis system of the crystal {[100], [010], [001]}, the second order bulk nonlinear susceptibility has only three non-vanishing, equal tensor components, $\chi^{(2)}_{ijk} = \chi^{(2)}_{jki} = \chi^{(2)}_{kij}$, $\{i,j,k\} \in \{x,y,z\}$ such that the nonlinear polarization of second harmonic generation (SHG) in contracted notation matrix form is given by $P_i(2\omega) = 4\epsilon_0 d_{14} E_k E_j$ and the nonlinear polarization of sum frequency generation (SFG) in contracted notation matrix form is given by: $P_i(2\omega) = 4\epsilon_0 d_{14}[\, E_j(\omega_1)E_k(\omega_2) + E_k(\omega_1)E_j(\omega_2)]$. For the top and bottom surface of GaAs, the crystal symmetry at the surface leads us to the following relations for the second order susceptibility. $P_x^{(2)} = 4\epsilon_0 d_{15}^s E_z E_x$, $P_y^{(2)} = 4\epsilon_0 d_{24}^s E_z E_y$ and $P_z^{(2)} = 4\epsilon_0[\, d_{31}^s E_x^2 + d_{32}^s E_y^2 + d_{33}^s E_z^2]$. Since the orientation of the array is at 45 degrees with the respective crystal axes, we have expressed the nonlinear polarizability in the laboratory frame by using Euler rotation matrices. The full expression of the second and third order nonlinear polarizabilities used in our full-wave simulation are given in supplementary note S1. We also run any additional simulations with different coefficient values of the third order nonlinear tensors while respecting the symmetry constraint imposed to confirm the fourfold symmetry character of $\chi^{(3)}$: THG.

**Author Contribution** S.D.G., I.B designed the study; S.D.G simulated the linear response, S.D.G, P.P.I measured the linear spectra, S.D.G characterized the nonlinear response, S.D.G, C.D., N.K. developed the nonlinear numerical model. S.D.G wrote the manuscript with inputs from all authors.

**Funding.** U.S. Department of Energy (BES 20-017574);

**Acknowledgments.** This work was supported by the U.S. Department of Energy, Office of Basic Energy Sciences, Division of Materials Sciences and Engineering and performed, in part, at the Center for Integrated Nanotechnologies, an Office of Science User Facility operated for the U.S. Department of Energy (DOE) Office of Science. Sandia National Laboratories is a multi-mission laboratory managed and operated by National Technology and Engineering Solutions of Sandia, LLC, a wholly owned subsidiary of Honeywell International, Inc., for the U.S. Department of Energy's National Nuclear Security Administration under contract DE-NA0003525. This paper describes objective technical results and analysis. Any subjective views or opinions that might be expressed in the paper do not necessarily represent the views of the U.S. Department of Energy or the United States Government.

**Disclosures.** The authors declare no conflict of interest

**Supplemental document**. See [insert link by publisher] for additional information on fabrication, optical apparatus, linear and nonlinear modeling.